\documentclass[useAMS,usenatbib]{mn2e}
\usepackage[dvips]{graphicx}
\usepackage{longtable}

\def\kms{\mbox{$\rm km~s^{-1}$}}

\def\deg{$^\circ$~}
\def\arcsec{^{\prime\prime}~}

\begin{document}

\title[The Low Surface Brightness Galaxy PGC~045080]
{The AGN and Gas Disk in the Low Surface Brightness Galaxy PGC~045080}

\author[Das et al.]{M.Das$^{1}$, N.Kantharia$^{2}$, S.Ramya$^{3}$, T.P.Prabhu$^{3}$, S.S.McGaugh$^{4}$,
S.N.Vogel$^{4}$\\
1.Raman Research Institute, Bangalore, 560080, India\\
2.National Centre for Radio Astrophysics, Tata Institute of Fundamental Research, Post Bag 3,  Ganeshkhind,
Post Bag 3, Ganeshkhind, Pune - 411007, India\\
3.Indian Institute of Astrophysics, Koramangala, Bangalore, India\\
4.Department of Astronomy, University of Maryland, College Park, MD 20742, USA\\
(E-mails:mousumi{@}rri.res.in)}

\date{Accepted.....; Received .....}


\maketitle


\begin{abstract}
We present radio observations and optical spectroscopy of the giant low surface brightness
(LSB) galaxy PGC~045080 (or 1300+0144). PGC~045080 is a moderately distant galaxy having a highly
inclined optical disk and massive HI gas content. Radio continuum observations of the galaxy
were carried out at 320~MHz, 610~MHz and 1.4~GHz. Continuum emission was
detected and mapped in the galaxy. The emission appears extended
over the inner disk at all three frequencies. At 1.4~GHz and 610~MHz it appears to have two
distinct lobes. We also did optical spectroscopy of the galaxy
nucleus; the spectrum did not show any strong emission lines associated with AGN activity 
but the presence of a weak AGN cannot be ruled out. Furthermore,
comparison of the H$\alpha$ flux and radio continuum at 1.4~GHz suggests that a significant
fraction of the emission is non-thermal in nature. Hence we conclude that a weak or hidden AGN 
may be present in PGC~045080.
The extended radio emission represents lobes/jets from the AGN. These
observations show that although LSB galaxies are metal poor and have very little star formation,
their centers can host significant AGN activity. We also mapped the HI gas
disk and velocity field in PGC~045080. The HI disk extends 
well beyond the optical disk and appears warped. In the HI intensity maps, the disk appears 
distinctly lopsided. The velocity field is disturbed on the
lopsided side of the disk but is fairly uniform in the other half. We derived the HI rotation curve
for the galaxy from the velocity field. The rotation curve has a flat rotation speed of $\sim190$\kms.
\end{abstract}

\section{INTRODUCTION}

Low surface brightness (LSB) galaxies have diffuse stellar disks, large HI disks and 
low metallicities (Impey \& Bothun 1997; Bothun et al. 1997). They are poor in
star formation which is usually localized to relatively small patches over their disks 
(McGaugh 1994; de Blok \& van der Hulst 1998; Kuzio de Naray et al. 2004).
Very few LSB galaxies have been detected
in molecular gas, which is a fairly reliable indicator of the star forming potential
in a galaxy ((O'Neil, Hofner \& Schinnerer 2000;
Matthews \& Gao 2001; O'Neil, Schinnerer \& Hofner 2003; Matthews et al. 2005;
Das et al. 2006). The combined low star formation rates and metallicities of these 
galaxies suggest that they are less
evolved compared to high surface brightness (HSB) galaxies (McGaugh \& Bothun 1994; de Blok et al. 1995). They
are in some sense more $"$primordial$"$ than their HSB counterparts. One of the reasons for
the lack of evolution could be the massive dark halos that dominate the disks of these 
galaxies (e.g. de Blok \& McGaugh 1997). Dark matter halos inhibit the growth of disk 
instabilities such as bars and spiral arms which can act as triggers of large scale star 
formation in galaxies and help propel their evolution (Mihos, McGaugh \& de Blok 1997).

Recent studies indicate that LSB galaxies
span a wide range of morphologies from the more populous dwarf spiral/irregular galaxies to the
large spirals like Malin~1 (McGaugh, Schombert \& Bothun 1995; Galaz et al. 2002). The large LSB
galaxies or the so called giants are relatively rarer than the smaller LSB galaxies (Bothun
et al. 1990; Sprayberry et al. 1995). Many giant LSB galaxies have large bulges associated with extended
diffuse stellar disks (Beijersbergen et al. 1999). Although their disks
are faint, they often show a distint spiral structure (e.g. UGC~6614; Pickering et al. 1997)
and in some rare cases are even barred (e.g. UGC~7321; Pohlen et al. 2003). 

The optical spectrum of many
LSB giants show emission lines characteristic of an active galactic nucleus 
(Sprayberry et al. 1993; Schombert 1998). This is suprising as AGNs are usually associated 
with optically bright 
galaxies that are undergoing significant star formation  (Ho, Philipenko \& Sargent 1997). 
At least $20$\% of LSB galaxies show spectroscopic signs of AGN activity (Impey, Burkholder 
\& Sprayberry 2001). However, detailed investigations of the AGN activity at wavelengths other than
optical have not been done. A handful of giant LSB galaxies are bright at radio frequencies  
in the NVSS and FIRST VLA surveys (e.g. 
UGC~1922 in O'Neil \& Schinnerer 2004) but deeper observations are lacking. Das et al. (2006) detected
a millimeter continuum source in UGC~6614; the emission is flat spectrum between 1.4 to 110~GHz and is 
due to the AGN in the galaxy. These results suggest that the AGN in LSB galaxies may have similar 
properties compared to HSB galaxies even though their evolutionary histories are very different.

We have undertaken a radio study of several giant LSB galaxies with the Giant Meterwave Radio Telescope
(GMRT) at metre wavelengths. This paper on PGC~045080 is the first in this study. We have also 
done optical spectroscopy of the nucleus using the Himalayan Chandra Telescope (HCT). 
In the following sections we present an overview of the galaxy, the HCT observations, the GMRT radio
observations and our results. Finally we discuss the implications of our observations. As 
PGC~045080 has a moderately high redshift (z=0.0409) we will adopt a
luminosity distance of $D_{L}=176.4~Mpc$ ($H_{0}=71~kms^{-1}~Mpc^{-1}$) which leads to a distance scale of 
0.86~kpc~arcsec$^{-1}$. 

\section{THE LSB GALAXY PGC~045080}

PGC~045080 is a fairly isolated giant LSB galaxy. The main properties of the galaxy are listed in Table 1. 
The galaxy is close to edge on, has a faint nucleus and very low surface brightness disk. An 
oval structure is visible in the galaxy center and may represent a weak bar or a bulge. 
Although highly inclined, spiral arms are clearly 
visible in the galaxy disk. The optical spectrum of PGC~045080 was observed by Sprayberry et al. (1995)
as part of a larger study of the nuclei of giant LSB galaxies. Although they found signatures of
AGN activity in several other LSB galaxies, they did not find AGN activity in
PGC~045080. However, their spectral resolution was rather poor and weak AGN activity may have been missed.

As in most LSB galaxies the neutral hydrogen (HI) content of the galaxy is
high compared to its stellar mass. Single dish observations indicate that the HI gas disk has a double horned 
profile typical of a rotating gas disk
(Sprayberry et al. 1995; Matthews et al. 2001). The  
HI mass is $\sim10^{10}~M_{\odot}$ for this galaxy (Impey, Burkholder \& Sprayberry 2001). 
However, neither the HI gas distribution nor disk kinematics in 4 have been mapped before. 

\section{GMRT RADIO OBSERVATIONS}
 
We observed 4 with the GMRT on four days from August 2005 to March 2006. The details of the observations
are listed in Table 2. The GMRT is an array of thirty antennas of 45m diameter each; 14 antennas are arranged in 
a central compact array and the remaining are distributed over a Y shaped configuration (Swarup et al. 1999;
Ananthakrishnan \& Rao 2002). Nearby radio source were used for phase calibration. The 
data was obtained in the native "lta" format, converted to FITS format and then analysed using AIPS 
\footnote{AIPS is
distributed by NRAO which is a facility of NSF and operated under cooperative agreement by
Associated Universities, Inc.}. 
    
Continuum observations were done at frequencies 1.4~GHz, 610~MHz and 325~MHz. 
Bad data on a single channel on the amplitude and bandpass calibrator were
iteratively edited and calibrated until satifactory gain solutions were obtained using standard tasks in 
AIPS. This was used to generate
bandpass solutions. The central 85 channels were bandpass calibrated and averaged to obtain the
continuum database. This was then imaged using IMAGR. Several nearby bright sources were used to self calibrate
the continuum images. A relatively bright point source
lies approximately $2^{\prime}$ to the north west of PGC~045080; the source is seen in the FIRST and NVSS VLA maps
and has a peak flux of $\sim$80~mJy. The presence of this bright source limited the sensitivity of the map 
close to the galaxy. Wide field imaging
was used at 610~MHz and 325~MHz; at 610~MHz the primary beam was divided into 9 facets and at 325~MHz it was
divided into 25 facets. At all frequencies both natural and uniform weighted maps of PGC~045080 were made.

For the HI line observations, the data were also gain and bandpass calibrated. The HI data cube was generated 
from the UV line data using a maximum UV distance of 40 Klambda. Since a bandwidth of 8~MHz was used for the 
observations, the resulting resolution was 62.5~KHz or 14.3\kms. The task MOMNT in AIPS was then used to generate 
moment maps from the data cube using a cutoff of 1.2~mJy/beam (i.e. 2$\sigma$). These maps reveal the
HI intensity distribution and the gas kinematics in the galaxy disk.
The velocity field was also used to derive the HI rotation curve; this was done using the progam rotcur which 
is part of the software toolbox NEMO (Teuben 2001). Rotcur uses the 
tilted ring analysis of Begeman (1989). The
HI disk was divided into annuli and the velocity averaged for each ring. Nyquist sampling was used to maximize
the number of annuli sampled. We used rotcur to find a dynamical center for the galaxy; it is  
offset by approximately $1\arcsec$ from the SDSS optical center of the galaxy. 

\section{RESULTS OF RADIO OBSERVATIONS}

Radio continuum emission from this galaxy was detected for the first time.  
In previous VLA observations at 1.4~GHz (NVSS maps; Condon et al. 1998), the galaxy falls below 
NVSS detection limits,
although there is a peak in the NVSS map at the 2$\sigma$ level. 
The galaxy is not detected in the higher
resolution FIRST survey (Becker, White \& Helfand 1995). However both surveys did not go very deep. 
Using the GMRT we have detected radio continuum emission
from the center of PGC~045080 at all three frequencies i.e. 1.4~GHz, 610~MHz and 325~MHz.
Apart from radio continuum observations, we have also mapped the HI gas distribution in the galaxy.
Previous HI observations were all single dish and did not give any informations about the gas 
morphology or kinematics. In the following paragraphs we discuss our results in more detail;
the flux estimates and beams have been summarised in Table 3.

\subsection{RADIO CONTINUUM OBSERVATIONS}

\noindent
{\bf (i)~1.4~GHz} The distribution of the 1.4~GHz radio continuum emission is shown in Figures~1 to 4. 
Figure~1 shows the high sensitivity map made with natural weighting. The resolution is rather poor
(beam$\sim8\arcsec$) but the extended flux distribution is clearly seen. There is a peak in the flux 
distribution that has a
S/N ratio of 8 and is offset from the nucleus of the galaxy by $1-2\arcsec$. The peak flux of 
the emission is $\sim0.71~mJy~beam^{-1}$ and it extends over a region $\sim10\arcsec-15\arcsec$ in 
size. The emission also has a southern spur like feature (Figure~2). This 
is clearer in the higher resolution map made with uniform weighting (Figure~3 and 4), where the
southern extension has a S/N~4 but does not not appear to be associated with the galaxy. At 610~MHz this
low level emission extending south of the galaxy falls below the 2$\sigma$ level. Hence it is 
unlikely to be nonthermal in nature and associated with a background galaxy. It is also unlikely 
to be due to star formation as there is no emission associated with it in the optical images (Sprayberry et al. 
1995); thus the emission extending to the south may not be real. 

At high resolution (beam$\sim4.3\arcsec$) in Figure~3 the 1.4~GHz  
emission breaks up into two main cores. The 
galaxy center lies closer to the brighter peak but does not lie 
exactly between the two peaks (Figure~4).
Hence from this map it is not clear whether the emission has a double lobe structure about the 
nucleus or not.  However, at both resolutions the 
emission lies along an axis which is slightly offset from the optical axis of the 
galaxy. This suggests that the continuum emission may not be due to star formation in the 
galaxy disk but instead may be associated with an AGN in the galaxy. If so, then the 
extended emission may  
be jets or lobes originating from the active nucleus. This is discussed in more detail in Section~7.

\noindent
{\bf (ii)~610~MHz}
Figure~5 shows the radio continuum emission from PGC~045080 at 610~MHz. This is the low resolution but 
high sensitivity map made with natural weighting. The emission has an extended structure as in the
1.4~GHz emission map (Figures 1 to 4). The emission peaks on one side of the galaxy and 
the peak has a S/N ratio of 6. The emission extends over $\sim15\arcsec$ and the peak flux
is $\sim0.88~mJy~beam^{-1}$ where the beam is $\sim10\arcsec$. 
In the higher resolution map made with uniform weighting, the extended emission breaks up into two lobes
about the center of the galaxy (Figure~6). Both high and low resolution maps suggest a double lobed 
morphology for the emission and the lobe east of the galaxy center is possibly the closer lobe 
as it appears brighter
in both the high and low resolution maps. 

\noindent
{\bf (iii)~325~MHz}
At 325~MHz the galaxy is detected in the radio continuum with a S/N of 4 but the map is noisy
and the resolution is poor. Hence 
not much can be said about the structure except that it is extended. The peak flux in the map is 
$\sim3.56~mJybeam^{-1}$ where the beam$\sim13\arcsec$.

\noindent
{\bf (iv)~Spectral Index}
The radio maps of PGC~045080 are too noisy to derive meaningful spectral index maps. So instead we 
used the peak flux in the 1.4~GHz and 325~MHz maps to calculate the spectral index. 
We obtained a value of $\alpha=-0.63\pm0.2$ where
$\alpha$ is defined as $S_{\nu}\propto\nu^{\alpha}$. Within error limits the spectral index is what is
expected from either star forming regions or jets associated with an active nucleus.

\subsection{HI GAS DISK}

Although the HI content of PGC~045080 had been studied earlier in single dish observations,
the gas distribution had not been studied. Hence this is the first HI map of this galaxy. 
The results of our GMRT HI line observations of PGC~045080 are discussed below. 

\noindent
{\bf (i)~HI disk size and mass:} The most striking feature
of the HI distribution in PGC~045080 is its extent compared to its stellar disk.
Figure~7 shows the HI intensity map of PGC~045080 and Figure~8 shows the intensity contours overlaid on the
SDSS R band optical image of the galaxy. The gas disk is about twice the size of the optical disk and 
possibly much thicker as
well. The optical disk has a diameter of approximately $40\arcsec$ whereas the gas disk has a diameter 
of $80\arcsec$. A
larger HI disk relative to the optical disk is typical in LSB galaxies (Pickering et al. 1997). 
But it is more clearly seen in this galaxy as it is nearly edge on. 
The total HI flux is $\sim0.86~Jy~kms^{-1}$ which leads to a 
HI gas mass of $6.3\times10^{9}$~M$_\odot$. This is approximately two thirds of the single dish HI flux observed
with the Arecibo telescope by Sprayberry et al. (1995) which is $9.6\times10^{9}$. The difference is 
probably because of the limitation in uv space coverage in our GMRT observations (although we had 2 tracks 
of data on this galaxy, only one data set was included in the channel map because data 
quality of the second days observation was poor). Assuming that the HI gas disk has a 
radius $\sim40\arcsec$, the mean
surface density of HI is $\Sigma\sim1.7~M_{\odot}pc^{-2}$. 
This is lower than that observed in other giant LSB galaxies; for example UGC~6614 has a corresponding 
value of $\Sigma\sim3.2~M_{\odot}pc^{-2}$ (Pickering et al. 1997). 

\noindent
{\bf (ii)~Lopsided HI distribution:}
The HI intensity distribution on either side of the galaxy nucleus is different and gives the galaxy a 
``lopsided" appearance. The peak HI intensity east of the nucleus in the disk is about 1.4 times the peak 
on the west. The lopsided nature of the HI disk peaks just at the edge of the optical disk ($\sim17\arcsec$).
One side of the disk is also more flared indicating that the HI disk is perhaps warped. A warped disk has also
been seen in a more prominent LSB galaxy, Malin~1 (Pickering et al. 1997). 

\noindent
{\bf (iii)~Minimum in HI distribution:}
There is
a dip in the HI intensity on brighter side of PGC~045080 near the galaxy center; the origin of the HI dip is 
not clear as there is no bright star formation in this region. However, it may however 
be related to the extended radio
continuum emission observed in the galaxy because the peak emission of the lobe east of galaxy center 
is close to the dip in HI intensity. Figure~9 shows the 610~MHz map contours overlaid on the HI map. Note that 
the brighter lobe that lies east of the galactic center coincides with the dip in the HI intensity. This 
indicates that there may have been some disk-jet interaction that 
resulted in some of the HI being evaporated away leaving behind a dip in the gas distribution.

\noindent
{\bf (iv)~HI velocity field:}
The intensity weighted velocity field or moment~1 map of HI disk is shown in Figure~10 The velocity contours
on one side of the disk (west of nucleus) look fairly typical of a rotating disk. But the velocity field on
the other side (i.e. the flared half of the disk) is not uniform. This again suggests that the HI disk
is warped or lopsided and hence the velocity field is non-symmetric about the galaxy center.

\noindent
{\bf (v)~HI Rotation Curve:} 
The intensity weighted velocity field was used to derive the HI rotation curve for the galaxy. As discussed
earlier in Section 5 we used rotur to determine a dynamical center for the galaxy and it is 
offset by approximately $1\arcsec$ from the SDSS center of the galaxy. We plotted the rotation curves
for the approaching and receding sides separately (Figure~11). The two curves are significantly different in the
inner $15\arcsec$ radius of the disk, but have similar speeds in the outer region. The two curves were averaged
to obtain the mean rotation curve for the galaxy (Figure~12). The curve is relatively flat
until a radius of approximately $35\arcsec$; beyond that it
starts rising. At radii larger than $\sim45\arcsec$ there is not much gas  and hence not enough points to
fit a reliable rotation curve.
The flat rotation speed for 4 is $\sim190$~\kms.

\section{HCT OPTICAL OBSERVATIONS}

The radio continuum observations discussed in Section 4.1 suggest the presence of an AGN in PGC~045080. In 
particular the 
double lobed structure seen in the 610~MHz map (Figure~6) is often observed in galaxies harboring radio loud AGNs.
To get a better understanding of the nuclear activity in PGC~045080 we did  
optical spectroscopy of the galaxy nucleus with the HCT, which is a 2m optical telescope located
in the Hanle valley in Ladakh. Long slit spectroscopic observations of the central bright nucleus were obtained
on 29th and 30th April, 2006. We used a 1671 slit (dimensions $1.92\arcsec\times11^{'}$) and a combinations
of two grisms, Grism~7 (3800 to 6840~$\AA$) which has a resolution of 1.45~A$^{o}$~pixel$^{-1}$ and Grism~8
(5800 to 8350~$\AA$) which has a resolution of 1.25~$\AA$~pixel$^{-1}$. The exposures in both the grisms
was one hour each on both nights. For wavelength calibration we observed Fe-Ar and Fe-Ne calibration lamps for
a few seconds each. Flux calibration was done by observing the standard star Feige~34.

The nuclear spectrum was extracted over 8.88" length along the slit.
Since the nights were not photometric, the continuum fluxes obtained for the two nights differed.
So we used the SDSS
spectrum of this galaxy as a secondary calibrator to scale the flux levels and fit the continuum shape.
Data reduction was done using IRAF \footnote{IRAF is distributed by the NOAO
which is operated by AURA under cooperative agreement with NSF.} with the task APALL in the SPECRED
package. The task FITPROFS was used to deblend the spectral lines and derive the total flux, equivalent widths
and FWHM of the lines.

\section{RESULTS OF OPTICAL SPECTROSCOPY}

The final spectrum is shown in Figure~13 and the details of the emission lines are listed in Table 4.
The spectrum has a narrow but prominent
H$\alpha$ line and strong [SII] and [NII] lines in emision. The emission
lines are superimposed on a stellar continuum and absorption line
spectrum. The H$\beta$ and H$\gamma$ lines are seen in absorption
suggesting significant star formation has taken place in recent past.
The signatures of older population are also evident making it difficult
to model the star-formation history. We could not identify any strong
emission lines indicative of AGN, such as broad H$\alpha$ and high
excitation lines such as [OIII], HeII and lines of Ne. The intermediate
excitation lines of [FeVI] 5158\AA\ and [FeVII] 6058\AA\ are detectable.
However, they can arise due to AGN as well as SN II in the post-starburst
phase. Thus the optical spectrum of the galaxy nucleus does not provide
an unambiguous signature of AGN in PGC~045080.

We also used the flux values of the emission lines to determine the position of PGC~045080 on the AGN
diagnostic diagrams (Baldwin et al. 1981). The diagnostic diagrams provide a very good tool to distinguish
between AGN and starburst emission in galaxies (Groves, Heckman, Kauffman 2006; Kewley et al. 2006).
Figure~14 shows the plot of log(O[III]/H$_{\beta}$) against log([NII]/H$_{\alpha}$). The solid line is
the limit for extreme starburst regions from Kewley et al. (2001) and the dashed line is
the lower limit between AGN and starburst nuclei from Kauffman et al. (2003). The region between these
two lines forms the composite region where the spectra may contain contributions from both AGN and starburst
emission. 
PGC~045080 lies on the the dashed line and is hence on the boundary between HII regions and AGN/starburst nuclei.
The low value of [OIII]/H$\beta$ excludes a Seyfert type nucleus, and at
most a weak LINER may contribute to the optical spectrum. To summarise, the emission line 
spectrum of the the nucleus
does not show any clear signatures of AGN activity. At most the AGN may be a weak LINER that is visible
at radio frequencies but not in the optical.

\section{IS THERE AN AGN IN PGC~045080?}

We have done radio observations and optical spectroscopy of the nucleus of PGC~045080.
Extended radio
continuum emission was detected from the galaxy at 1.4~GHz, 610~MHz and 325~MHz. The emission
morphology suggests that it is not
associated with star formation in the disk but may instead be due to AGN activity.
Optical spectroscopy
of the nucleus was done to see if the spectrum had emission lines characteristic of an AGN.
However as discussed in Section~6 we did not see any outstanding emission lines
in the optical spectrum that confirms the presence of an AGN in PGC~045080. The H$\alpha$ line is prominent
but narrow and the O[III] line is relatively weak. Considering its position on the diagnostic diagram we conclude
that at most there may be a weak AGN in the galaxy and it may host LINER type activity. However another
approach is to compare the expected radio continuum from H$\alpha$ flux with that actually observed at 1.4~GHz.
We used only the H$\alpha$ flux from the nuclear region over a slit dimension of $1.92\arcsec\times8.88\arcsec$.
Assuming that the strong H$\alpha$ line arises from star formation, we corrected the spectrum for reddening
using the standard balmer decrement method used by Kong et al. (2002), wherein the intrinsic theoretical
H$\alpha$/H$\beta$ is asumed to be 2.87 for starburst galaxies. The E(B-V) for this galaxy is then 0.153~magnitude.
After correcting for reddening, the nuclear H$\alpha$ luminosity amounts to $2.09\times10^{40}$~erg~s$^{-1}$.
This is an upper limit for the expected H$\alpha$ flux from star formation. The resulting infrared flux is 
$\sim 1.18$~erg~s$^{-1}$cm$^{-2}$ (Kennicutt
1998) and the expected radio continuum emission at 1.4~GHz from nuclear star formation is 
0.23~mJy (Condon et a. 2002).
We then used the GMRT radio continuum map (Figure~1) to determine the 1.4~GHz flux from PGC~045080 over a 
region similar to that of the optical slit used in
the HCT observations. This was done to compare as precisely as possible the expected radio flux from star formation 
with that actually observed in the GMRT maps. The integrated flux over the nuclear region is 0.54~mJy which 
is more than two times that
expected from star formation. Hence a significant fraction of the
radio continuum emission from PGC~045080 is non-thermal in nature and is most likely
due to an AGN in the galaxy. Also, there is no significant H$\alpha$ flux detected from the disk in the spectrum; the
H$\alpha$ emission arises mainly from the nucleus. (This is clear when we compared the H$\alpha$ emission from the slit
integrated over the disk with that over the nucleus). 
This further supports the idea that the extended radio continuum
is non-thermal in nature and is associated with an AGN in the galaxy.

\section{DISCUSSION}

We have done radio observations and optical spectroscopy of the nucleus of PGC~045080.
Extended radio
continuum emission was detected from the galaxy at 1.4~GHz, 610~MHz and 325~MHz. 
We also mapped the extended HI gas distribution and velocity field in the galaxy.
In the following section we discuss the implications of our observations.  

\noindent
{\bf (i) Radio Jets in PGC~045080:} If the radio continuum is from an AGN in the galaxy, the morphology 
suggests that we are detecting only the radio jets or lobes and not the core. This is not unusual and is seen 
in more powerful radio galaxies as well (e.g. NGC~3801, Das et al. 2005). The core emission is self absorbed and 
becomes visible only at higher frequencies. From the higher resolution maps at 1.4~GHz and 610~MHz it is clear
that the radio lobe lying east of the nucleus is brighter than that on the west. The galaxy center is also 
marginally closer to the brighter lobe. This may indicate that the lobe to the east is closer to us than that on
the western side. However since the resolution is poor, details of the galaxy and radio lobe morphology are 
difficult to understand. Higher resolution observations are required to fully understand the radio jet/lobe
morphology. 

\noindent
{\bf (ii) Radio Power in Jets :} Assuming that the extended emission is mainly non-thermal in nature, we can
calculate the radio power in the lobes and compare it with other radio sources.
Using the formula for equipartition of energy between magnetic fields and particle
energies (Longair 1983) the total energy in synchroton emission in the lobes (assuming a frequency of 325~MHz)
is $\sim1.5\times10^{55}~erg$. This is similar to the energy stored in compact
cores or jets in Seyfert galaxies and also in FR~I type radio sources. Hence these observations
indicate that though LSB galaxies have a low star formation rate, they can host AGN activity comparable to 
bright spirals and even radio galaxies. How AGN activity developed in these poorly evolved, low surface 
brightness galaxies is something we need to understand. 

\noindent
{\bf (iii) Lopsided HI gas disk :} PGC~045080 is clearly asymmetric in the HI maps and to some 
degree in the optical image as well. Spiral
galaxies have been found to be lopsided both in HI (Baldwin et al. 1980; Richter \& Sancisi 1994) and in 
the near-infrared (Rix \& Zaritsky 1995; Zaritsky \& Rix 1997). The origin for the asymmetry can 
be due to ram pressure in a cluster enviroment or due to interactions with nearby 
galaxies (e.g. Angiras et al. 2006). It can also be due to an off center galaxy potential 
(Jog 1997; Levine \& Sparke 1998)
or slow gas accretion (Bournaud et al. 2005). Lopsidedness is also more commonly seen in late 
type galaxies, which are generally more halo dominated in their inner disks than early type 
galaxies (Matthews et al. 1998).
Although PGC~045080 is a relatively isolated
galaxy, it appears distinctly lopsided in the HI moment maps (Figures 7, 8 and 10). 
It also appears slightly asymmetric in the R band SDSS 
optical image, although it is difficult to tell as the galaxy is nearly edge on. 
Considering that PGC~045080 is a LSB galaxy and possibly has a massive dark halo, the most probable 
explanation for its lopsidedness is that its disk and halo potentials are off center with respect to 
each other (Noordermeer et al. 2001). Its also likely that a significant fraction of LSB galaxies
are lopsided in their HI distribution as seen in the LSB dwarf galaxy DD0~9 (Swaters et al. 1999).

\section{CONCLUSIONS}

We have done an optical and radio study of the relatively distant LSB galaxy PGC~045080. The main results 
of our work are listed below.

\noindent
{\bf (i) Optical Spectrum :} The emission lines in the optical spectrum do not indicate the presence of
strong AGN activity (i.e. a Seyfert type nucleus). At most a LINER type nucleus may be present.  

\noindent
{\bf (ii) Extended Radio Continuum Emission :} We detected extended radio continuum emission from the 
galaxy at 1.4~GHz, 610~MHz and 320~MHz. The spectral index is $\alpha=-0.63\pm0.2$. The 
radio morphology suggests that the emission is not due to star formation but instead represents lobes or
jets associated with an AGN in the galaxy. 

\noindent
{\bf (iii) AGN and radio jets in the galaxy :} We compared the radio flux expected from the H$\alpha$
emission to that actually observed at 1.4~GHz. We found that a significant fraction of the emission 
is non-thermal in nature and probably due to AGN activity in the galaxy. 
If so, then the extended radio continuum observed at 1.4~GHz and 610~MHz are radio lobes associated 
with the AGN. Such structures have not been seen before in LSB galaxies. It suggests that these galaxies
may harbour AGN that have radio properties similar to Seyfert galaxies. 

\noindent
{\bf (iv) HI morphology and kinematics :} The HI disk extends out to approximately twice the optical 
size of the galaxy. The HI morphology is asymmetric w.r.t the galaxy center and appears thicker on 
the eastern side compared to the other half. The disk kinematics are also less regular in the flared 
or warped side compared to the more regularly rotating western half. The rotation curve yields a flat
rotation velocity of $\sim190\kms$.

\noindent
{\bf (v) Lopsided HI disk :} The HI gas disk appears to be lopsided in the moment maps. As
PGC~045080 is a fairly isolated galaxy, we think that the asymmetry may be due to an offset 
between the center of the disk and dark matter halo in the galaxy.

\section{ACKNOWLEDGMENTS}

We are grateful to the members of the GMRT staff for their help in the
radio observations. The GMRT is operated by the National Centre for Radio Astrophysics of 
the Tata Institute of Fundamental Research.
We thank the staff of IAO, Hanle and CREST, Hosakote, that made these obervations possible. 
The facilities at IAO and CREST are operated by the Indian Institute of Astrophysics, Bangalore.
This paper has used an SDSS image; funding for the SDSS and SDSS-II has been 
provided by the Alfred P. Sloan Foundation, the Participating Institutions, NSF,
the U.S. Department of Energy, NASA, the Japanese Monbukagakusho, the Max Planck Society, 
and the Higher Education Funding Council for England. This research has
made use of the NASA/IPAC Infrared Science Archive, which is operated
by the JPL, California Institute of Technology,
under contract with NASA.

\clearpage

{\begin{table}
\caption{Parameters of PGC~045080}
\noindent
\begin{tabular}{lcc}
\hline
\hline
Parameter & Value & Comment \\
\hline
Galaxy Type & Sc(f) & NED \\
Galaxy Position (RA, DEC) & 13:03:16.0 +01:28:07 & 2MASS \\
Other Names &   1300+0144, CGCG 015-059 &  \\
Velocity \& Redshift & 12,264~\kms, 0.040908 & NED \\
Luminosity Distance & 176.4~Mpc & \\
Linear Distance Scale & 0.86~kpc/arcsec & \\
Disk Inclination & 71\deg & Sprayberry et al. 1995 \\
Disk Optical Size ($D_{25}$) & 0.55' & Sprayberry et al. 1995 \\
Galaxy Position Angle (P.A.) & 85\deg & 2MASS \\
\hline
\hline
\end{tabular}
\label{Table 1}
\end{table}
}

\vspace{5mm}

{\begin{table}
\caption{Details of GMRT Observations}
\begin{tabular}{lcccc}
\hline
\hline
  Name of Record     & HI line   &  1.4~GHz        &  610~MHz       &  325~MHz \\
\hline
Date of Observation  & 19, 20 August 2005 &  19 August 2005 &  17 March 2006 & 12 December 2005 \\
Central Frequency    & 1364.4~MHz     &  1364.4~MHz     &  604.5~MHz     & 317.6~MHz   \\
Bandwidth            &  5.4~MHz       &   5.4~MHz        &  8.7~MHz       &  12.5~MHz \\
Phase Calibrator     &  1347+122      & 1347+122       &  1419+064      &  1419+064 \\
Amplitude Calibrator &  3C147, 3C286  & 3C147, 3C286    &  3C147         &  3C147, 3C286 \\
Bandpass Calibrator  &  3C147, 3C286  & 3C147, 3C286   &  3C147         &  1419+064  \\
Onsource Time        &  10 hours      & 6.5 hours      &  5.5 hours     &  3 hours  \\
Band Center (HI Line) & 12,264~\kms   & .....          &  ......        &  ......  \\
Spectral Resolution  &  14.3~\kms (62.5~kHz) & .....   & ......   &  ......  \\
Beam Size (Natural Weight) & $14.7\arcsec\times12.0\arcsec$  & $8.4\arcsec\times7.6\arcsec$ & $11.7\arcsec\times9.0\arcsec$ & $14.5\arcsec\times11.6\arcsec$ \\
\hline
\hline
\end{tabular}
\label{Table 2}
\end{table}
}

\vspace{5mm}

{\begin{table}
\caption{Results of Radio Continuum Observations of PGC~045080}
\begin{tabular}{lccc}
\hline
\hline
  Name of Record     &  1.4~GHz     &  610~MHz &  325~MHz \\
\hline
Peak Flux ($mJy~beam^{-1}$) & 0.71  & 0.88  &  3.6 \\
Noise                       & 0.08  & 0.15  &  0.9 \\
Beam Size (Natural Weight) & $8.4\arcsec\times7.6\arcsec$ & $11.7\arcsec\times9.0\arcsec$ & $14.5\arcsec\times11.6\arcsec$ \\
\hline
\hline
\end{tabular}
\label{Table 3}
\end{table}
}

\vspace{5mm}

{\begin{table}
\caption{Results of Optical Observations}
\begin{tabular}{lcccc}
\hline
\hline
Emission Line & Central Wavelength (A$^{o}$) & Flux (erg~cm$^{2}$~s$^{-1}$) & Equivalent Width (A$^{o}$) & FWHM (A$^{o}$) \\
\hline
H$_{\beta}$ (emission)   &   4862.178    &     1.707E-15    &   3.818    &  11.23  \\
H$_{\beta}$ (absorption) &   4860.973    &    -2.33E-15$^{a}$ &   5.206    &  20.97  \\
O[III]                   &   5005.22     &     3.092E-16    &   0.6974   &   8.127 \\
Fe[VI]                   &   5158.461    &     3.642E-16    &   0.8163   &   7.648  \\
Mg (absorption)          &   5176.959    &     -1.76E-16$^{a}$ &   3.927     &  16.22   \\
Fe[VII]                  &   6084.591    &      2.387E-16    &  0.5076  &   6.632  \\
O[I]                     &   6300.673    &      3.231E-16    &  0.7001  &   11.2   \\
N[II]                    &   6548.056    &      8.724E-16    &  1.917   &   12.9   \\
H$_{\alpha}$             &   6564.066    &      5.708E-15    &  12.81   &   10.0   \\
N[II]                    &   6584.366    &      2.947E-15    &  6.797   &   11.14  \\
S[II]                    &   6717.033    &      8.190E-16    &  2.617   &   10.92  \\
S[II]                    &   6732.403    &      6.105E-16    &  1.964   &    9.506 \\
\hline
\hline
\end{tabular}
\begin{flushleft}
(a)~The negative sign implies an absorption line.\\
\end{flushleft}
\label{Table 4}
\end{table}
}

\clearpage 

\begin{figure}
\includegraphics[width=90mm,height=90mm]{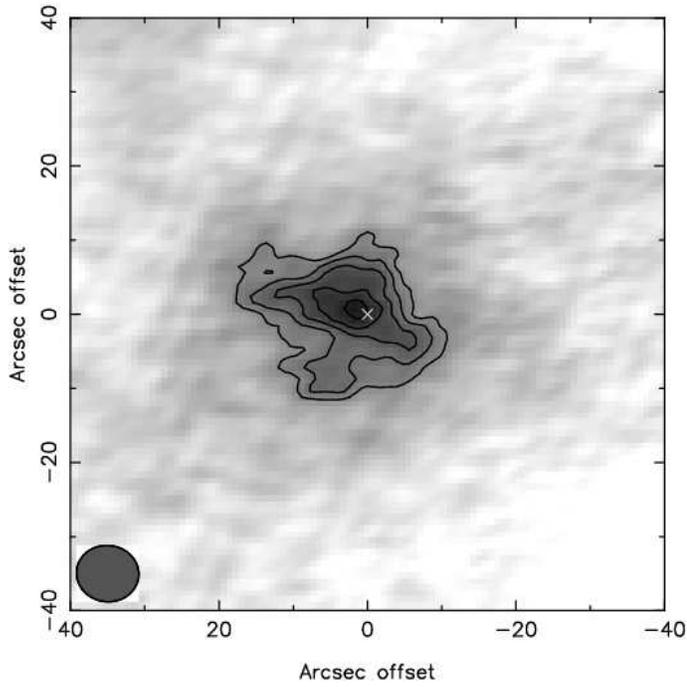}
\caption{Low resolution map of the 1.4~GHz continuum emission from PGC~045080. Natural weighting
has been used; the contours are 4, 5, 6, 7 and 8 times the map noise which is 0.085~mJy/beam.
The beam is $8.4\arcsec\times7.6\arcsec$ and the SDSS galaxy center is marked with a cross.}
\end{figure}

\vspace{10mm}

\begin{figure}
\includegraphics[width=90mm,height=90mm]{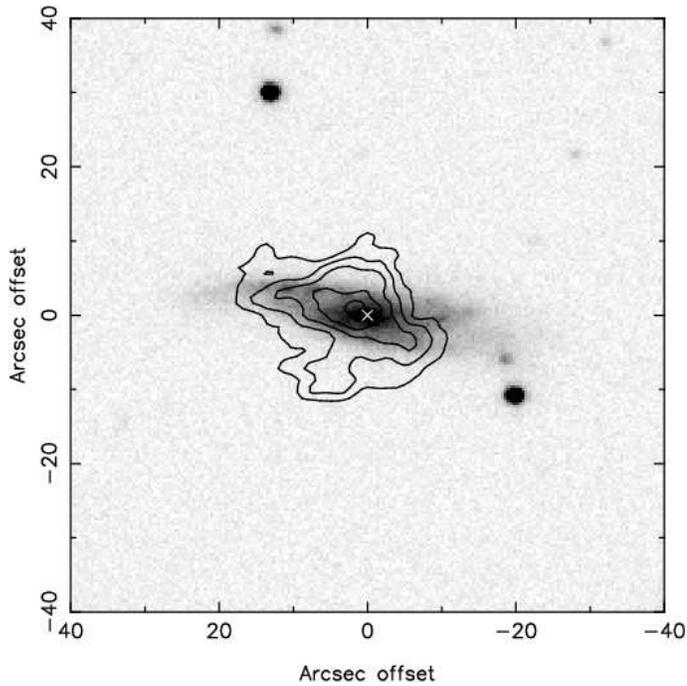}
\caption{SDSS R band image of PGC~045080 with the radio contours of 1.4~GHz emission at low resolution
from Figure 2a overlaid. Note that the extended radio emission to the south lies outside the galaxy and
is not associated with the galaxy; its origin is unclear.}
\end{figure}

\clearpage

\begin{figure}
\includegraphics[width=90mm,height=90mm]{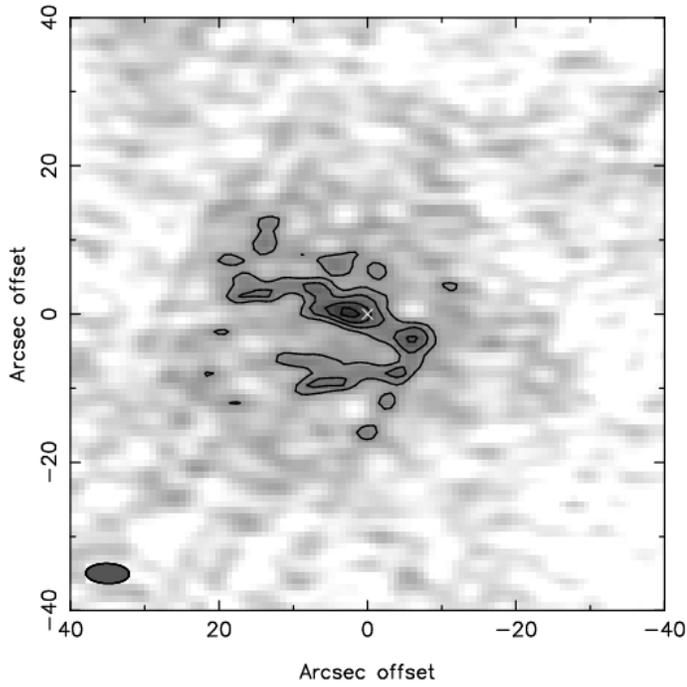}
\caption{Map of the 1.4~GHz continuum emission from PGC~045080. Uniform weighting has been used
to make the map and the contours are 3, 4, 5 and 6 times the map noise which is 0.088~mJy/beam
where the beam is $5.9\arcsec\times2.7\arcsec$.
The galaxy center from SDSS is marked with a cross.}
\end{figure}

\vspace{10mm}

\begin{figure}
\includegraphics[width=90mm,height=90mm]{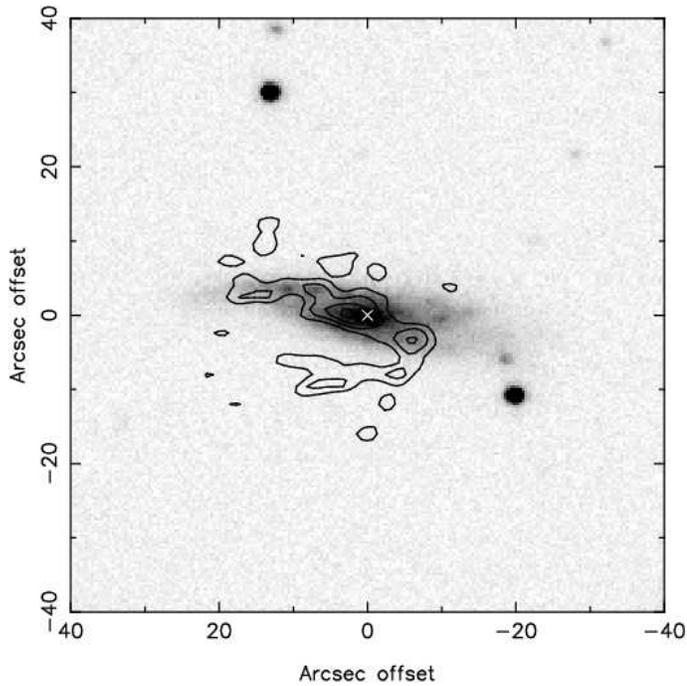}
\caption{SDSS R band image of PGC~045080 with the radio contours of 1.4~GHz emission from Figure~1
overlaid. Note that the P.A. of the radio axis is significantly different from the optical axis.}
\end{figure}

\clearpage

\begin{figure}
\includegraphics[width=90mm,height=90mm]{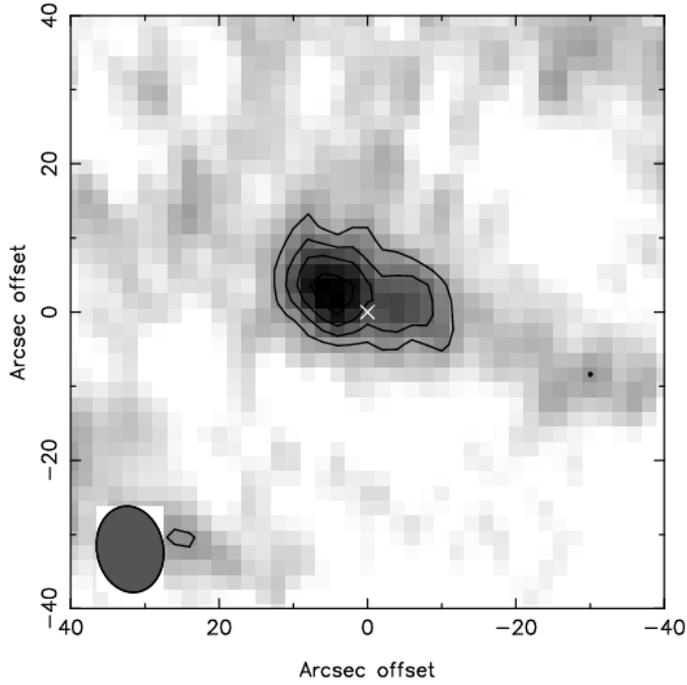}
\caption{The 610~MHz continuum emission map of PGC~045080; natural weighting has been used so the resolution
is poor but the sensitivity is relatively higher. The contours are 2, 3, 4 and 5 times the noise level
which is 0.16~mJy/beam where the beam is $11.7\arcsec\times9.0\arcsec$. The galaxy is detected with a 
S/N ratio of 6.}
\end{figure}

\vspace{10mm}

\begin{figure}
\includegraphics[width=90mm,height=90mm,angle=270]{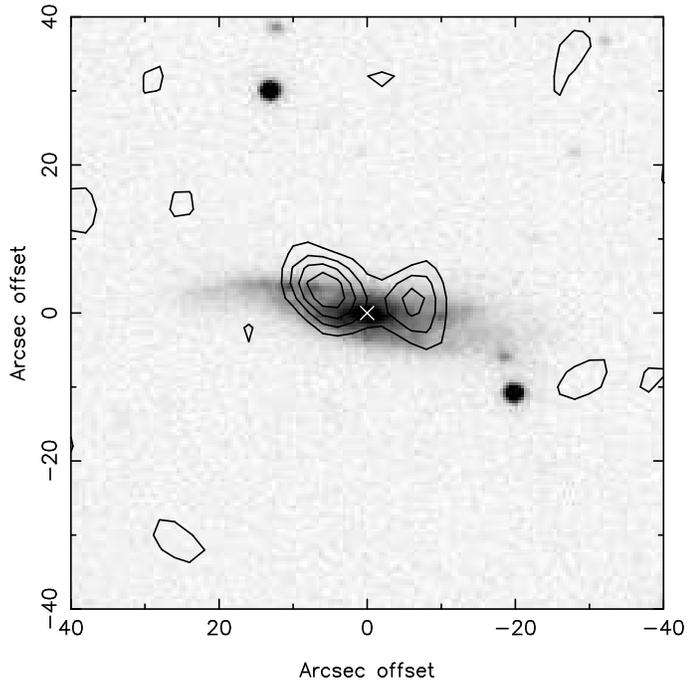}
\caption{SDSS R band image of PGC~045080 with the radio contours of 610~MHz emission map overlaid. The 
610~MHz image was made using uniform weighting to achieve better resolution (beam=$7.6\arcsec\times6.5\arcsec$). 
The 
contours are at 2, 3, 4 and 5$\sigma$ levels where $\sigma$=0.12~mJy~beam$^{-1}$. }
\end{figure}

\vspace{10mm}

\clearpage

\begin{figure}
\includegraphics[width=90mm,height=90mm,angle=270]{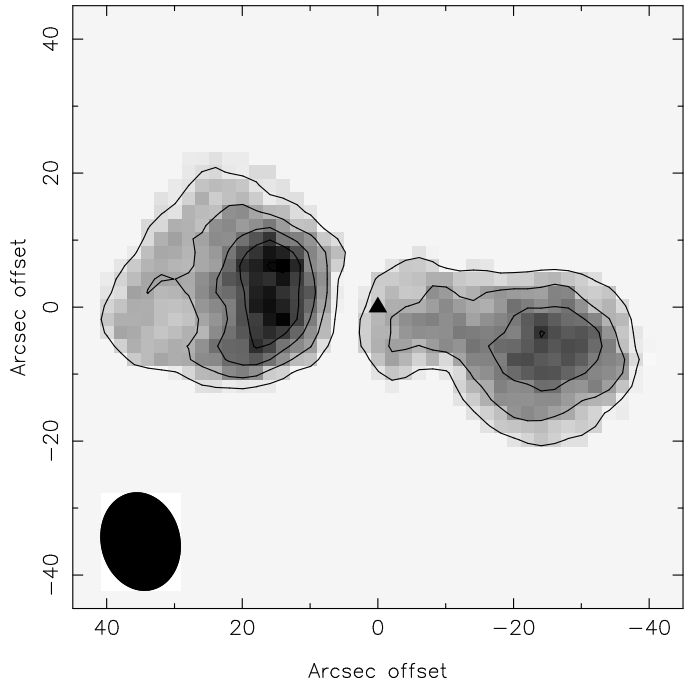}
\caption{Map of the HI intensity distribution in PGC~045080 observed with the GMRT. The contours 
are at 0.1, 0.3, 0.5, 0.7 times the peak emsission which is 266~mJy~beam$^{-1}$~\kms and the 
beam is $14.7\arcsec\times12.0\arcsec$. Note the gas morphology appears different on either side
of the galaxy center. }
\end{figure}

\vspace{10mm}

\begin{figure}
\caption{Contours of the HI intensity overlaid on the SDSS R band image of PGC~045080. The HI extent
is nearly twice that of the visible optical disk and appears thicker as well. }
\end{figure}

\clearpage

\begin{figure}
\includegraphics[width=90mm,height=90mm,angle=270]{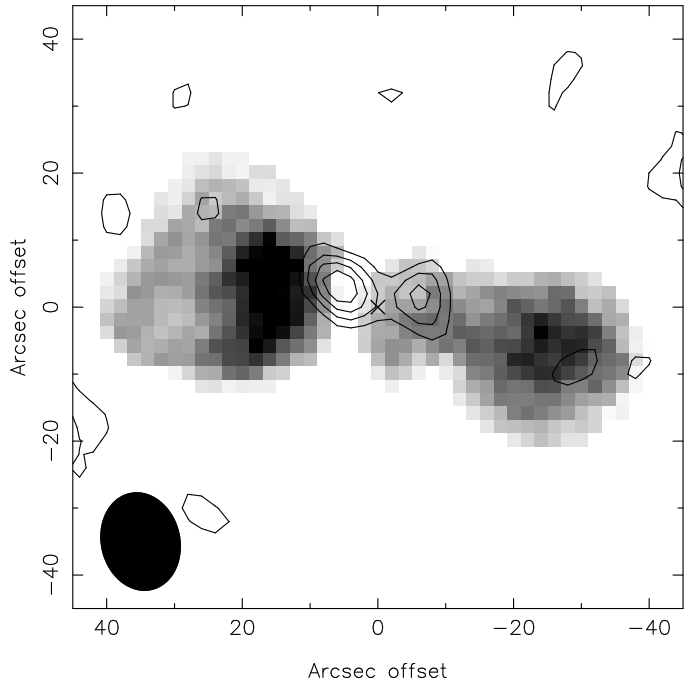}
\caption{Moment zero or HI intensity map of PGC~045080 with the contours of the high resolution  610~MHz 
emission overlaid. The brighter lobe of the continuum emission coincides with the dip in the HI 
distribution suggesting that there may have been some jet-cloud interaction in the disk of the galaxy. The
contours of the 610~MHz emission is the same as that in Figure~6.
}
\end{figure}

\vspace{10mm}

\begin{figure}
\caption{Figure shows the HI velocity field field of the galaxy PGC~045080. The velocity contours are
spaced 20$~\kms$ apart and are from -170 to 170~$\kms$, where the central velocity is taken to be the 
systemic velocity of the galaxy i.e. 12,264~$\kms$. The contours in green represent the approaching 
side while (east of the nucleus) while those in black represent the receding side.
}
\end{figure}

\clearpage

\begin{figure}
\includegraphics[width=60mm,height=100mm,angle=270]{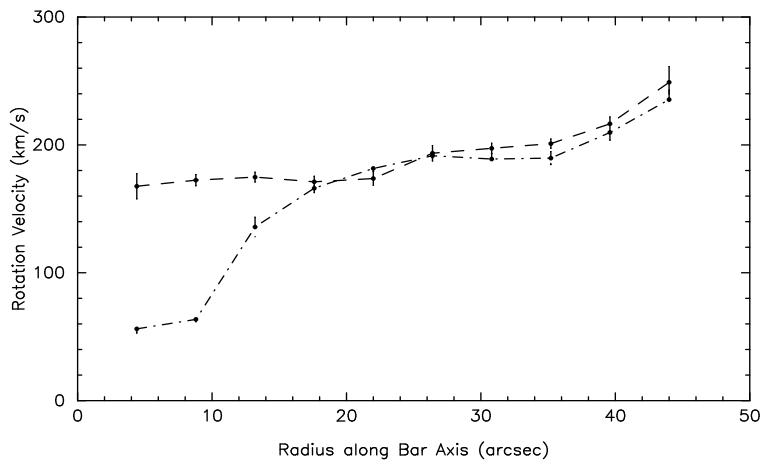}
\caption{The rotation velocities for both sides of PGC~045080 derived from the HI moment map (beam is 
$9.2\arcsec\times8.4\arcsec$. Nyquist sampling was used and velocities averaged over annuli of width
$\sim8.8\arcsec$ The approaching side is indicated with a dashed
curve and the receding with a dash-dot-dash line. The two sides agree beyond $20\arcsec$ approximately
which is the approximate radial extent of the stellat disk.}
\end{figure}

\vspace{30mm}

\begin{figure}
\includegraphics[width=60mm,height=100mm,angle=270]{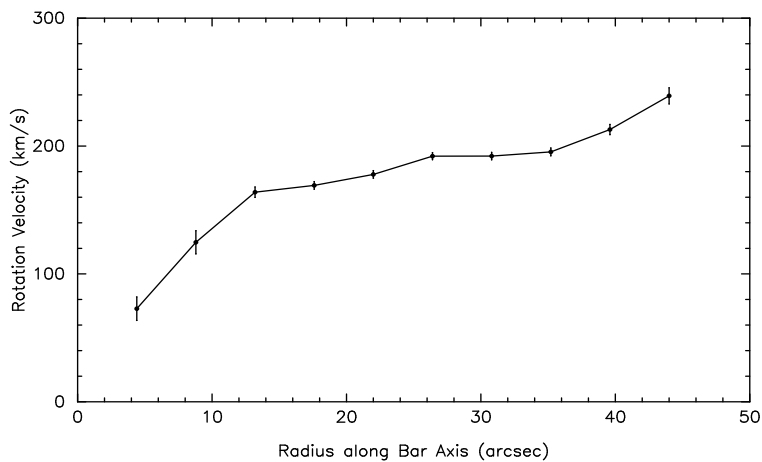}
\caption{The total rotation curve of the galaxy after summing up the contributions from both the 
approaching and receding sides. The error bar in for each velocity bin is plotted as well. The mean
flat rotation velocity is $\sim190\kms$.}
\end{figure}

\vspace{30mm}

\clearpage

\begin{figure}
\caption{The optical spectrum of PGC~045080. The slit was centered on the galaxy center and
aligned along the major axis. The main emission lines, H$\beta$, O[III], H$\alpha$, N[II] doublet and
S[II] are marked on the plot. The Y axis is the flux in erg~cm$^{-2}$~s$^{-1}$ and the X axis is
the wavelength in Angstrom.
}
\end{figure}

\vspace{10mm}

\begin{figure}
\includegraphics[width=90mm,height=80mm]{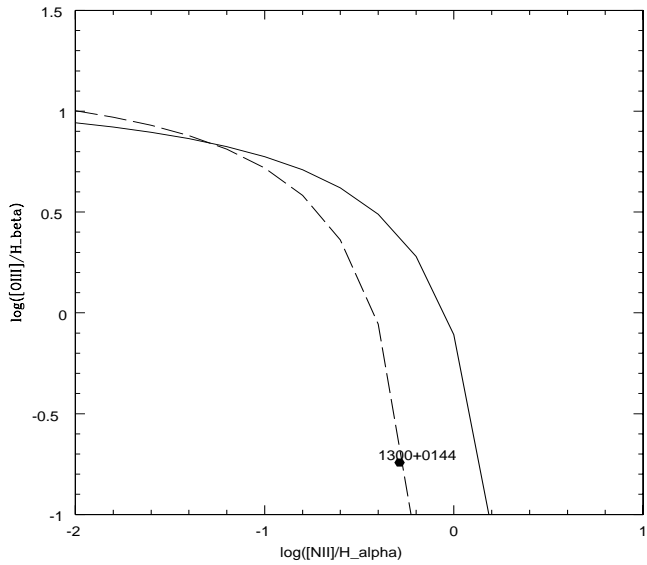}
\caption{Diagnostic diagram of the emission lines from the nucleus of PGC~045080. The solid line gives
the limit of extreme starburst (Kewley et al. 2001) and the dashed line is the limit for HII regions
from Kauffman et al. (2003). PGC~045080 lies on the border of the dashed line in the lower half of the
plot. Hence is it may have only a weak AGN, perhaps a LINER type nucleus.
}
\end{figure}

\end{document}